\documentclass[%
superscriptaddress,
preprint,
 amsmath,amssymb,
 aps,
 prl,
]{revtex4-2}

\usepackage{graphicx}
\usepackage{dcolumn}
\usepackage{bm}
\usepackage{comment}
\usepackage{units}
\usepackage[T1]{fontenc}
\usepackage{xcolor}
\usepackage{xr}

\usepackage{lineno}

\usepackage{hyperref}

\begin{document}

\title{Surface-Sensitive Raman Scattering by Transferable Nanoporous Plasmonic Membranes}

\author{Roman M. Wyss}
\affiliation{Institut f\"ur Physik und IRIS Adlershof, Humboldt-Universit\"at zu Berlin, 12489 Berlin, Germany}
\affiliation{Soft Materials, Department of Materials, ETH Z\"urich, 8093 Zürich, Switzerland}

\author{G\"unter Kewes}
\affiliation{Institut f\"ur Physik und IRIS Adlershof, Humboldt-Universit\"at zu Berlin, 12489 Berlin, Germany}

\author{Martin Frimmer}
\affiliation{Photonics Lab, ETH Z\"urich, 8093 Zürich, Switzerland}

\author{Karl-Philipp Schlichting}
\affiliation{Laboratory of Thermodynamics in Emerging Technologies Department of Mechanical and Process Engineering, ETH Zürich, 8092 Zürich, Switzerland}

\author{Markus Parzefall}
\affiliation{Photonics Lab, ETH Z\"urich, 8093 Zürich, Switzerland}

\author{Eric Bonvin}
\affiliation{Photonics Lab, ETH Z\"urich, 8093 Zürich, Switzerland}

\author{Martin F. Sarott}
\affiliation{Laboratory for Multifunctional Ferroic Materials, Department of Materials, ETH Zürich, 8093 Z\"urich, Switzerland}

\author{Morgan Trassin}
\affiliation{Laboratory for Multifunctional Ferroic Materials, Department of Materials, ETH Zürich, 8093 Z\"urich, Switzerland}

\author{Lala Habibova}
\affiliation{Institut f\"ur Physik und IRIS Adlershof, Humboldt-Universit\"at zu Berlin, 12489 Berlin, Germany}

\author{Giorgia Marcelli}
\affiliation{Institut f\"ur Physik und IRIS Adlershof, Humboldt-Universit\"at zu Berlin, 12489 Berlin, Germany}

\author{Jan Vermant}
\affiliation{Soft Materials, Department of Materials, ETH Z\"urich, 8093 Zürich, Switzerland}

\author{Lukas Novotny}
\affiliation{Photonics Lab, ETH Z\"urich, 8093 Zürich, Switzerland}

\author{Mads C. Weber}
\affiliation{Institut des Mol\'ecules et Matériaux du Mans, UMR 6283 CNRS, Le Mans Université, 72085 Le Mans, France}

\author{Sebastian Heeg}
\email[Correspondence email address: ]{sebastian.heeg@physik.hu-berlin.de}
\affiliation{Institut f\"ur Physik und IRIS Adlershof, Humboldt-Universit\"at zu Berlin, 12489 Berlin, Germany}

\date{
\today
}

\begin{abstract}
Raman spectroscopy is a powerful technique to characterize materials. It probes non-destructively chemical composition, crystallinity, defects, strain and coupling phenomena. However, the Raman response of surfaces or thin films is often weak and obscured by dominant bulk signals. Here we overcome this limitation by placing a transferable porous gold membrane (PAuM) on top of the surface of interest. Slot-like nanopores in the membrane act as plasmonic slot antennas and enhance the Raman response of the surface or thin film underneath. Simultaneously, the PAuM suppresses the penetration of the excitation laser into the bulk, efficiently blocking the bulk Raman signal. Using graphene as a model surface, we show that these two simultaneous effects lead to an increase in the surface-to-bulk Raman signal ratio by three orders of magnitude. We find that \unit[90]{\%} of the Raman enhancement occurs within the top \unit[2.5]{nm} of the material, demonstrating truly surface-sensitive Raman scattering. To validate our approach, we analyze the surface of a LaNiO$_3$ thin film. We observe a Raman mode splitting for the LaNiO$_3$ surface-layer, which is spectroscopic evidence that the surface structure differs from the bulk. This result underpins that PAuM give direct access to Raman signatures of surfaces and their structural properties.
\end{abstract}

\maketitle

\section{Keywords}

\noindent Raman spectroscopy; surface; surface-sensitive Raman scattering; plasmonic nanopore; complex oxide thin film.

\section{Introduction}

Raman spectroscopy, the inelastic scattering of light by vibrations or phonons, is a widespread analytical tool to study and characterize materials \cite{pandey2021overview}. Owing to its versatility, simplicity and specificity, Raman spectroscopy is used in material science \cite{das2011raman}, i.e. to study phase transitions \cite{Zhang:2005a,ThuNguyen:2020a}, catalysis \cite{hess2021new} or novel 2D materials \cite{paillet2018graphene}, or even in pharmaceutics and biomedical diagnostics \cite{eberhardt2015advantages}. In principle, Raman spectroscopy is ideal to study the structure of surfaces, since their atomic registry differs from the bulk of the material and may additionally be modified by terminations. This leads to changes in the frequency of the Raman active vibrations or to peak-splitting as a result of a change insymmetry~\cite{Sarycheva:2020a,Pfisterer:2019a,Hayazawa:2007a}. However, the study of surfaces and thin films by Raman spectroscopy is notoriously difficult as light typically penetrates several micrometers into the material. The overall Raman response is, therefore, dominated by the bulk, while the Raman signals of the surface are orders of magnitude weaker and mostly go undetected. Hence, obtaining Raman signals of thin films requires a minimal thickness of several tens to hundreds of nanometers. Raman signatures of surfaces are often not observed at all \cite{ding2016nanostructure, gasparov2014thin}. 

One way to address the issue of vanishing surface- or thin film Raman signals is ultraviolet (UV) Raman spectroscopy. Here, a UV laser excites the sample instead of a laser in the visible or near infrared range. UV-Raman takes advantage of the shallow penetration-depth of the UV-light into many materials \cite{Tenne2006, Kreisel2012} and is not impeded by autofluorescence effects. However, the penetration of the laser in the material is still in the order of hundreds of nanometers, and can only be reduced further for materials with suitable band gaps~\cite{Kreisel2012}. Moreover, the low damage threshold of many materials to UV light limits the application of UV Raman \cite{Karim:2020a}. Probing surface and thin-film Raman signals has also been addressed by mathematical decomposition of a large stack of spectra. To do so, multiple spectra of a sample are measured under varying conditions. An example can be altering the laser focal point with respect the the sample surface. Subsequently, a statistical analysis allows to decompose the spectra into substrate/bulk and surface/thin film contributions \cite{Schober2020}. However, this method requires an already detectable signal of the surface or the thin film. Furthermore, the measurement times are often beyond practical use. 

A general strategy to enhance Raman signals is plasmon-enhanced Raman scattering (PERS)~\cite{RN956,Novotny:2006a}. Here, the enhancement in PERS occurs in the vicinity of metallic nanostructures and arises from the near-fields of localized surface plasmon resonances in the metal. Particularly striking enhancement occurs in a nanoscale gap between two metal nanostructures, also referred to as plasmonic hotspot. This configuration enables the detection of molecules adsorbed at the hotspot down to the single molecule level, embodying surface-enhanced Raman scattering (SERS)~\cite{RN952, RN955, SHARMA201216,Heeg:2020a}. Even though SERS can be realized with a large number of different nanoparticle geometries, its use to enhance the Raman signals of a surface or thin film is limited: there is no geometry that efficiently interfaces a plasmonic hotspot between two metallic structures with a flat and extended surface or film. Tip-enhanced Raman spectroscopy (TERS), where a plasmonic hotspot at the apex of a metal tip scans over a surface, partially solves this problem~\cite{STOCKLE2000131,zhang2007single}. The enhancement, however, occurs only at one spot and is weaker than for gap type geometries. The largest drawback of TERS is that bulk Raman signals are recorded together with the TERS signal. On top, TERS remains a complex and challenging technique, such that its use to probe surfaces or thin films is rarely reported. 

Overall, the ideal plasmonic structure to study surfaces and thin films with Raman spectroscopy consists of a flat gap type plasmonic hotspot with simultaneous bulk Raman signal suppression. Plasmonic nanoslots, rectangular nanoscale voids in a thin metallic film (i.e. nanoporous membrane), fulfill these conditions. Upon resonant excitation, these slots act as plasmonic slot antennas that harbour localized and enhanced near-fields, which rapidly decay outside the pore within few nanometers. Placed on a material, the slot’s near-fields interact primarily with the material surface. Away from the nanopore the metallic membrane reflects incident fields and bulk Raman signals, which effectively suppresses the bulk Raman signal. Recently, transferable and easy-to-manufacture porous gold membranes (PAuM) with nanoscale pores acting as plasmonic slot antennas were introduced~\cite{Wyss:2022a}. It was shown that individual pores feature local Raman enhancement factors up to $10^4$ to $10^5$ and sustain high excitation powers (\unit[$10^6$]{W cm$^{-2}$}), which makes them the ideal plasmonic structure to study surfaces and thin films with Raman spectroscopy.

Here, we use porous gold membranes to enhance the surface Raman signal and to simultaneously suppress the bulk signal of the sample under investigation. Using wavelength-dependent Raman spectroscopy, we show that PAuM enhance the surface-to-bulk Raman signal ratio by up to three orders of magnitude. Combining experiment and simulation, we reveal that the enhancement decays exponentially in the material such that \unit[90]{\%} of the enhanced signal occurs within the top \unit[2.5]{nm}. Hence, our approach enables highly surface-sensitive Raman spectroscopy for weak or bulk-obscured Raman signals. We directly apply this technique to study the surface of a \unit[20]{nm} LaNiO$_3$ thin film. We find Raman signatures of the surface that differ from the bulk of the \unit[20]{nm} film, in line with theoretical predictions and experimental observations using scanning tunneling microscope (STM) \cite{Fowlie2017}. Our work, therefore, underscores the power of PAuM-supported Raman spectroscopy of surfaces and thin films. 

\vfill

\section{Results}
Our paper is structured as follows: First, we introduce the PAuM manufacturing and its working principle. Second, we demonstrate the surface enhancement and bulk suppression of Raman signals by a wavelength-dependent study with graphene as a model surface. Third, we unravel the depth dependence of the Raman response in experiment and simulation. To do so, we probe graphene sheets buried at various depths from the surface. Finally, we showcase the use of PAuM to enhance the Raman response of a \unit[20]{nm} LaNiO$_3$ thin film.

\subsection{Manufacturing and Working Principle of $\text{PAuM}$}
\begin{figure}[!ht]
    \centering
    \includegraphics[width=13.6cm]{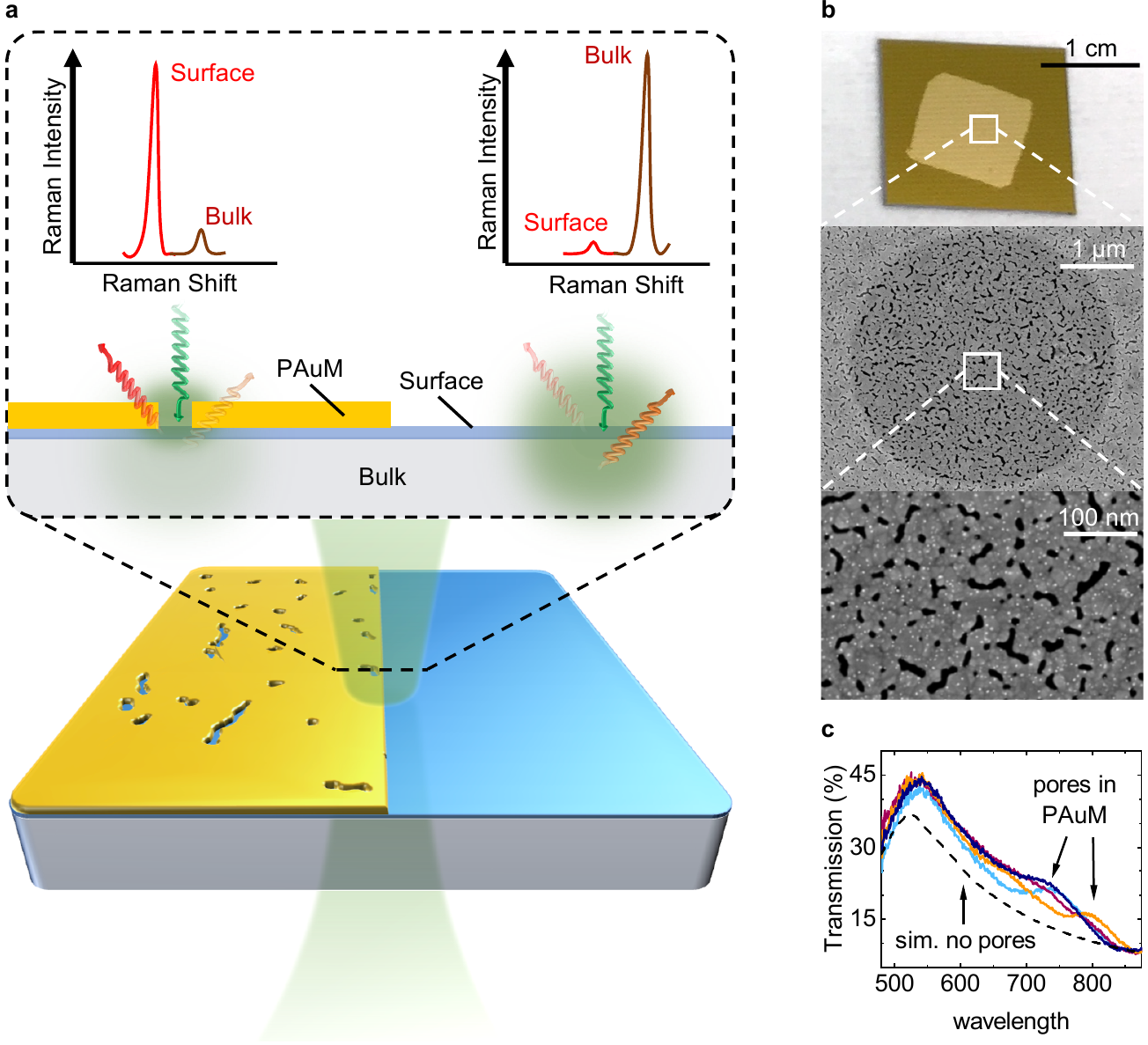}
    \caption{\textbf{Principle of surface-sensitive Raman scattering enabled by PAuM.} a) Nanopores in the gold membrane enhance the  Raman signal of the surface while suppressing the bulk Raman signals. b) Photographic image (top) of a \unit[20]{nm} PAuM transferred on a Si/Si$_3$N$_4$ partially suspended over \unit[4]{$\mu$m} holes (See Methods), forming freestanding membrane-like structures, as shown in the Scanning electron microscope (SEM) image (middle). The nanopores are visible as dark irregular, slot-like and circular features (bottom). c) Optical transmission of a freestanding PAuM measured as function of wavelength. The dashed line corresponds to a simulated non-porous Au film with \unit[20]{nm} thickness. The plasmonic resonances of the nanopores give rise to the increased transmission in the experimental data from \unit[650]{nm} to \unit[850]{nm} compared to the simulation of a non-porous film.}
    \label{fig:Introduction}
\end{figure}

Figure~\ref{fig:Introduction} (a) illustrates the key idea of this work: Without PAuM, a laser penetrates several multiples of its wavelength into the material, limited by absorption and focal depth. The Raman scattered signal, therefore, originates primarily from within the bulk. The surface Raman signal remains weak or non-detectable due to the vanishing small scattering volume of the surface. In contrast, using PAuM, the surface Raman signal is drastically enhanced compared the bulk Raman signal. This results from two simultaneous effects: 1) local plasmonic enhancement within the metallic nanopores and 2) the suppression of the laser penetration into the bulk and of residual bulk Raman signal reaching the detection pathway. Our non-continuous membranes are formed by evaporation of a \unit[20]{nm} gold film on a SiO$_2$/Si wafer. The membrane is subsequently transferred onto the sample \cite{Wyss:2022a} (see Methods and Supporting Information S1).

To characterize PAuM, we transfer a \unit[1]{cm$^2$}-sized membrane on a Si/Si$_3$N$_4$ chip bearing arrays of \unit[4]{$\mu$m} holes (Methods) as shown in the optical microscope image in Fig.~\ref{fig:Introduction}(b), top. The scanning electron microscope (SEM) images Fig.~\ref{fig:Introduction}(b), middle and lower panel, reveal that the PAuM spans over the circular hole as a freestanding, mechanically stable membrane \cite{Wyss:2022a}. The pores in the PAuM are visible in the SEM images as dark spots and come in various shapes and sizes. The majority of pores is round or slot-like and below \unit[100]{nm} in size (See Supporting Information S1). A recent study demonstrated that the nanopores in the PAuM act as plasmonic nanoslot antennas~\cite{Wyss:2022a}. The nanopores harbour intense light fields upon excitation at their plasmonic resonance. The energy of the plasmonic resonance depends one the shape and aspect ratio of the individual pores. The highest field enhancement occurs for narrow slot-like pores with an excitation polarized perpendicular to the pores' long axis, see Supporting Information and Ref.~\cite{Wyss:2022a} for an extended discussion. 

In Fig.~\ref{fig:Introduction} (c), we compare the optical transmission of freestanding \unit[20]{nm} PAuM to the transmission of a simulated gold film without nanopores of the same thickness. The general shape of the optical transmission is in good agreement with the simulated results (dashed line). However, deviations occur from \unit[650]{nm} to \unit[850]{nm} with an increased transmission probability. Such increased transmission is expected when the nanopores of PAuM are resonantly excited and act as nano-antennas~\cite{Park:2018a}. The colored lines correspond to individual transmission measurements at different sample positions, and their varying deviations from the simulated trend indicate the varying geometries of the nanopores. We conclude that Raman enhancement can be expected in the spectral range from \unit[650]{nm} to \unit[850]{nm} in agreement with our previous study~\cite{Wyss:2022a}. The random nature of pore geometries in both - dimension and orientation - provides a substantial number of pores that will resonantly couple to an incident laser light with arbitrary polarization and wavelength.

\subsection{Measurement of Surface Enhancement Using Graphene Probes}

Graphene is ideal as a model material to test and quantify surface-sensitive Raman scattering as it can mimic a surface or thin film due to its two-dimensional nature~\cite{Heeg:2012a,Heeg:2013a,Wasserroth:2018a}. Furthermore, the Raman spectrum of graphene is well understood and the main Raman features are intense and independent of excitation wavelength as well as polarization~\cite{Ferrari:2013a}. Any dependence of the Raman features of graphene interfaced with our porous Au membrane can hence be attributed entirely to nanopore interaction.  

\begin{figure}[!t]
    \centering
    \includegraphics[width=17cm]{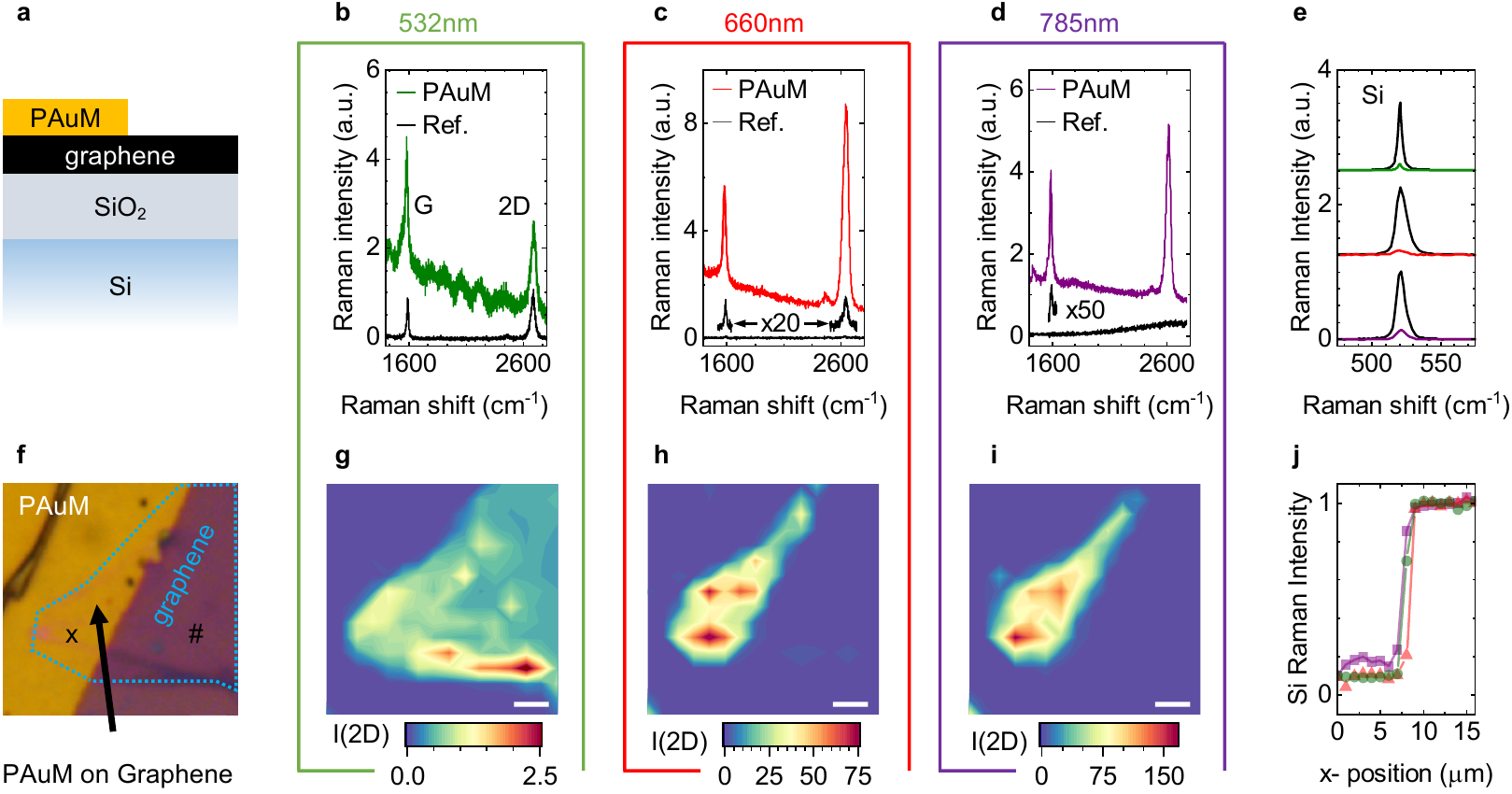}
    \caption{\textbf{Graphene as model surface to probe surface-sensitive Raman scattering by nanoporous Au membranes} a) Sample schematic where a graphene flake on a Si/SiO$_2$ substrate is partially covered by a PAuM. b-d) Raman spectra of the bare graphene (black, reference) compared to spectra from graphene plus PAuM for (b, green) \unit[532]{nm}, (c, red) \unit[660]{nm}, and \unit[785]{nm} (d, purple) excitation. The background in (b-d) stems from gold luminescence with an modulation due to etaloning for \unit[532]{nm}. e) 1st order Silicon Raman peak with PAuM for \unit[532]{nm} (green), \unit[660]{nm} (red), and \unit[785]{nm} (purple) normalized and compared to the corresponding reference Raman spectra (black) without PAuM. The spectra for each wavelength are offset for clarity. f) Microscope image of graphene flake (dashed turquoise) partially covered by PAuM (yellow). $\times$ (with PAuM) and $\#$ (reference) mark the locations of the spectra shown (b)-(e). g-i) Spatial Raman maps of the graphene 2D mode for (g) \unit[532]{nm}, (h) \unit[660]{nm}, and (i) \unit[785]{nm} excitation. For each excitation wavelength, the intensity is normalized to the 2D intensity of bare graphene reference with interference effect taken into account.  j) Spatial profile of the normalized silicon Raman intensity along a horizontal line through the locations x and \# as indicated in (f) for all three excitation wavelengths. The scale bar in (g-j) is $\unit[2]{\mu m}$.}
    \label{FIG:SSR_Exp}
\end{figure}

To probe the interaction and enhancement of PAuM with a surface, we place a graphene sheet, acting as an test surface, on a flat Si/SiO$_2$ substrate, see Supporting Information S2. A PAuM is transferred on top of this model system (see Methods) to cover the graphene sheet partially as sketched in Fig.~\ref{FIG:SSR_Exp}(a). In this way, we can probe the two effects that contribute to surface sensitive Raman scattering: First, the surface enhancement by the plasmonic nanopores of the PAuM via the graphene Raman signal and second, the bulk-signal suppression by monitoring the Raman signal of the Si substrate with and without the membrane. Figure~\ref{FIG:SSR_Exp}(f) shows a light microscope image of the PAuM-graphene-stack on a substrate. The PAuM appears as the yellow region, and monolayer graphene flake is marked by the dotted line. As can be seen in the image, the graphene flake is partially covered by the PAuM. 

First, we compare individual Raman spectra of PAuM-covered and uncovered graphene for three excitation wavelengths in Fig.~\ref{FIG:SSR_Exp}(b-d). The uncovered graphene serves as reference and the corresponding reference spectra are shown in black for each excitation wavelength, where $\times$ and $\#$ in Fig.~\ref{FIG:SSR_Exp}(f) mark the measurement positions. For \unit[532]{nm} excitation, Fig.~\ref{FIG:SSR_Exp}(b), the G- and 2D Raman modes of graphene show comparable intensities for PAuM-covered (green) and uncovered graphene (black). This behavior is expected since the nanopores in the membrane are not in resonance with the \unit[532]{nm} laser excitation~\cite{Wyss:2022a}. The 2D-to-G intensity ratio further confirms that the graphene in these measurements is indeed a monolayer ~\cite{Graf:2007a,Ferrari:2006a}. 

For $\unit[660]{nm}$ excitation, the Raman spectrum for PAuM-covered graphene (red ) is substantially enhanced when compared with uncovered graphene (black) in Fig.~\ref{FIG:SSR_Exp}(c). The enhancement is even more pronounced for $\unit[785]{nm}$ excitation, Fig.~\ref{FIG:SSR_Exp}(d). In this wavelength range, plasmonic enhancement from the pores comes into play, in good agreement with our transmission data, Fig.~\ref{fig:Introduction}(c), and previous works~\cite{Wyss:2022a}. For a quantitative analysis of the enhancement, we take care of potential reflection effects at the graphene-Si/SiO$_2$ interface~\cite{Yoon:2009a}. After excluding interference effects (Supporting Information), we find enhancement factors between $33$ and $165$ for the G and 2D modes (Table~\ref{TAB:SSR_factorss}). Note that the 2D-mode intensity for bare graphene and \unit[785]{nm} is below the noise level. We therefore use the noise level to approximate the 2D-mode intensity. 

The spatial Raman maps shown in Fig.~\ref{FIG:SSR_Exp}(g-i) trace the intensity of the 2D Raman mode of graphene normalized to the uncovered graphene reference for \unit[532]{nm}, \unit[660]{nm} and \unit[785]{nm} excitation, respectively. Interference effects are accounted for in all maps. The Raman map in Fig.~\ref{FIG:SSR_Exp}(g) confirms the negligible effect of the PAuM on the Raman response for \unit[532]{nm} excitation. In contrast, the Raman maps for $\unit[660]{nm}$ and $\unit[785]{nm}$ excitation show the striking enhancement by the PAuM. Clearly, the enhancement only occurs in areas of PAuM-covered graphene. Local variations in the enhancements reflect the random distribution of the plasmonic nanopores in the membrane with respect to geometry and orientation. 

\begin{table}[pb]
    \centering
    \caption{Enhancement factors, bulk suppression and Surface-enhancement $\times$ bulk-suppression as figure of merit for the overall increase for \unit[660]{nm} and  \unit[765]{nm} excitations.}
    
    \begin{tabular}{l|c|c|c|c}
$\lambda$  &    Enh (G)	& Enh (2D) &  Bulk suppression & Surface-enhancement $\times$ bulk-suppression\\
        \hline
    \unit[660]{nm}	  & $72$  &  $69$  & $10$ & $690$ to $720$\\
    \unit[785]{nm}	  & $33$  &  $165$ & $6.6$ &$220$ to $1100$ \\
 \end{tabular}
  
    \label{TAB:SSR_factorss}
 \end{table}

Second, we demonstrate the suppression of the bulk/substrate  Raman signal by the PAuM. To do so, we make use of the silicon substrate \unit[300]{nm} below the PAuM, where the silicon Raman signals are not enhanced. We compare the 1$^{st}$ order Raman peak of silicon at $\unit[521]{cm^{-1}}$ with (colored) and without a PAuM (black) for our three excitation wavelengths in Fig.~\ref{FIG:SSR_Exp}(e). The spectra indicate a clear suppression of the silicon Raman signal with PAuM for all wavelengths. This agrees with the attenuation expected for the incoming laser light and the Raman scattered light for a non-porous membrane of comparable thickness, see Fig.~\ref{FIG:SSR_Exp}(c). A line scan across the edge of the PAuM (Fig.~\ref{FIG:SSR_Exp}(j)), reveals a constant silicon Raman signal for all wavelengths on either side of the edge. The experimentally observed suppression by the PAuM amounts to a factor of 10 for $\unit[532]{nm}$ and $\unit[660]{nm}$, and 6.6 for $\unit[785]{nm}$ see Table~\ref{TAB:SSR_factorss}. Since the Si Raman signal without PAuM used as reference is affected by interference, the bulk Raman signal suppression a factor 5 to 8 higher then the experimentally obtained values, see Supporting Information S4. This brings the suppression closer to the values expected from our transmission experiments, see Fig.~\ref{fig:Introduction}(c). The exact magnitude of bulk Raman signal suppression depends on the exact geometry of the sample investigated. We therefore consider it most adequate to provide the experimental values as a lower bound for bulk Raman signal suppression by our PAuM. The product of surface Raman enhancement and bulk-suppression, which is the primary figure of merit for surface-sensitive Raman scattering as suggested here, then amounts to values between 220 and 1100, see Table~\ref{TAB:SSR_factorss}. 

\subsection{Depth Dependence of Raman Enhancement by PAuM}

\begin{figure}[pt]
    \centering
     \includegraphics[width=8cm]{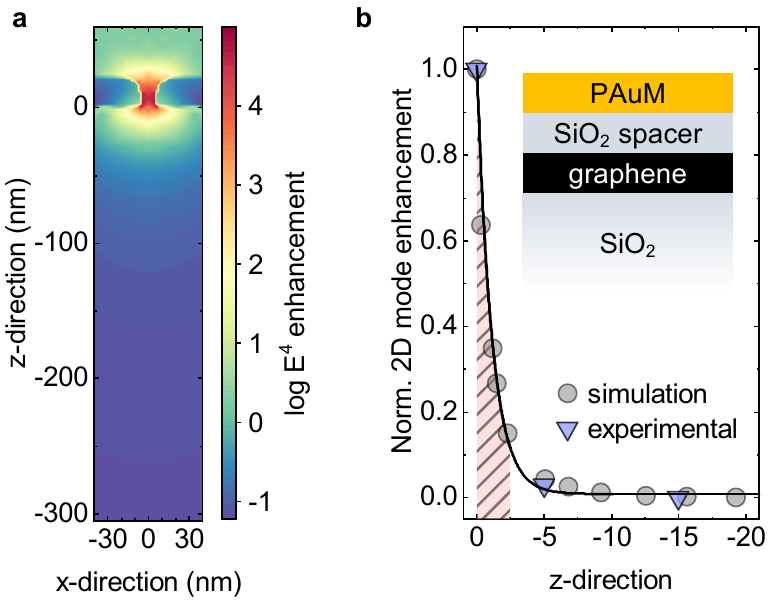}
    \caption{\textbf{Enhancement of Raman signal as a function of distance from the surface} a) Simulated E$^4$ enhancement (log scale) in the $xz$ plane for a \unit[22]{nm} thick gold membrane with a 10x68 nm slot on SiO$_2$. b) The experimental (triangle) and simulated (circles) enhancement for the graphene 2D Raman mode are shown as a function of SiO$_2$ spacer thickness between the porous Au membrane and graphene. The black line is an exponential fit to the simulated values. The dashed area marks the volume within which \unit[90]{\%} of the total Raman enhancement occurs. The simulated enhancement is normalized to the value at $z=\unit[-0.35]{nm}$, which is equivalent to the thickness of single layer graphene.}
    \label{FIG:Raman_Depth}
\end{figure}

Next, we investigate the effective Raman enhancement of the material below the PAuM as function of depth by simulation and experiment.  Since nanopores act as slot antennas~\cite{Wyss:2022a}, we simulate the Raman enhancement by a prototypical plasmonic nanoslot (\unit[10]{nm} $\times$ \unit[68]{nm}) similar in shape and size to nanopores found in the PAuM using a finite element solver JCMsuite (version 5.2.0) for Maxwell’s equations. We assume a  wavelength of \unit[660]{nm} for excitation and \unit[800]{nm}  for the Raman scattered light of the graphene 2D-mode (see methods and Supporting Information S5). Figure \ref{FIG:Raman_Depth}(a) depicts the simulated enhancement in the xz-plane along the nanoslot's short axis (x-direction). We then extract the average enhancement in the entire xy-plane from our simulation as function of depth $z$. To obtain the average, we do not only consider the areas directly under the nanoslots but also the area around it. This includes an area 15 times larger than the area of the slot and accounts for the fact that each nanopores is surrounded by continuous gold membrane segment, see Fig.~\ref{fig:Introduction}(b). We plot the average Raman enhancement of the graphene 2D-mode \textit{versus} $z$ in Fig.~\ref{FIG:Raman_Depth}(b) and find that the enhancement drops sharply with an increasing distance from the nanoslot. The decay is described by an exponential function $e^{z/\tau}$ with $\tau=\unit[1.1]{nm}$. This means that $90\%$ of the total Raman enhancement occurs within the first \unit[2.5]{nm} below the nanoslot.

In the next step, we probe the field enhancement as function of distance from our PAuM experimentally. 
To do so, we sputter \unit[5]{nm} and \unit[15]{nm} SiO$_2$ spacers on graphene and subsequently transfer PAuM on top as sketched in Fig.~\ref{FIG:Raman_Depth}(b), see Supporting Information S6-S8. We plot the enhancement of the graphene 2D-mode with and without the two different spacers  together with the simulation data in Fig.~\ref{FIG:Raman_Depth}(b). We find that the experimental enhancement is in excellent agreement with the simulation. This finding confirms that the PAuM-enhanced Raman spectroscopy allows for truly surface-sensitive Raman scattering, with an effective enhancement depth of less than \unit[5]{nm}.

The good agreement between our simulation and the experimental data for graphene allows us to infer more general properties of the surface enhancement provided by our PAuM. Extended simulations indicate an equally fast decay with distance from the PAuM for the entire relevant Raman frequency range of few cm$^{-1}$ to \unit[2500]{cm$^{-1}$} (Supporting Information S5). Furthermore, we find the strongest enhancement and the sharpest enhancement decay for nanopores with plasmonic resonances having high quality-factors. High quality-factors are inherent to narrow pores, i.e., gap features $< \unit[10]{nm}$~\cite{Wyss:2022a}. Wider pores with lower aspect ratios feature lower quality-factors, which leads to a reduced enhancement and greater decay constants $\tau\unit[\sim 5]{nm}$, equivalent to $90\%$ of the total Raman enhancement within the first \unit[11.5]{nm}. The approach to achieve the best surface-to-bulk enhancement ratio of the Raman signal is therefore to identify the highest Raman enhancement within a spatial map (see for example Fig.~\ref{FIG:SSR_Exp}) and evaluate its spectrum. 

\subsection{Application of surface-sensitive Raman scattering by PAuM: Probing the structural properties of an oxide thin-film surface}

In the next step, we further elucidate the use of the porous gold membranes for surface investigations. We choose a complex oxide, a LaNiO$_3$ thin film on LaAlO$_3$, as showcase material. Complex oxides are particularly suitable because their physical properties are highly sensitive to structural distortions~\citep{Lilienblum2015a, Giraldo2021, Liu2013a}. LaNiO$_3$ illustrates this well, as it is one of the few conducting perovskite-type materials and therefore an important electrode material for perovskite-type heterostructures~\citep{Catalano2018}. However, its conductivity is thickness-dependent, where below a thickness of 3 unit cells, LaNiO$_3$ even 
exhibits insulating behavior. These conductivity changes are attributed to structural inhomogeneities in the thin films. More specifically, the bulk and the surface of a LaNiO$_3$ film show different degrees of structural distortions~\citep{Fowlie2017}. 
A previous Raman study of LaNiO3 thin films confirmed the structural changes with film thickness~\citep{Schober2020}. However, these Raman spectroscopic measurements would only give information about the average structures and did not distinguish between surface, bulk or heterointerface. Here, using PAuM-enhanced Raman spectroscopy, we specifically target the surface structure of LaNiO$_3$ thin films to observe structural distortions directly. 

\begin{figure}[pt]
\centering
\includegraphics[scale=1]{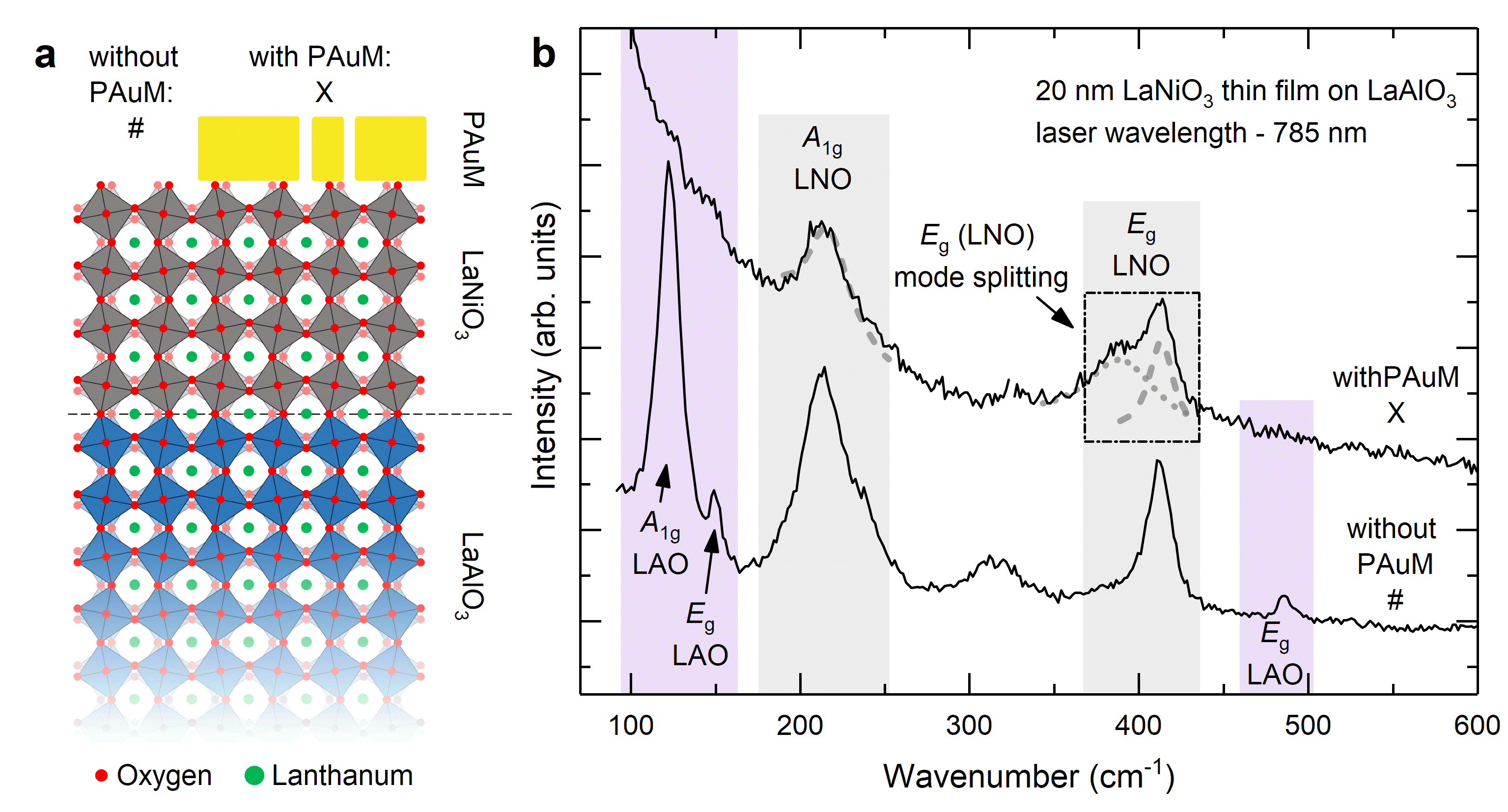}
\caption{\textbf{Surface Raman scattering of a LaNiO$_3$ thin film.} a) Sketch of the perovskite-type LaNiO$_3$ thin film on a (100)$_{\text{pc}}$-oriented LaAlO$_3$ substrate. The compound is partially covered by a PAuM. $\times$ and $\#$ indicate the measurement positions of the spectra in (b) with and without PAuM, respectively. b) The Raman spectrum of a LaNiO$_3$ thin film on LaAlO$_3$ with PAuM and without PAuM are depicted. Important Raman bands have been highlighted for visual guidance. The formation of a shoulder peak at \unit[389]{cm$^{-1}$} to the main peak at \unit[413]{cm$^{-1}$} is indicative of the distinct structural difference of the surface layer compared to the bulk of the LaNiO$_3$ thin film.}
\label{fig:LNO1}
\end{figure}

A \unit[20]{nm}-LaNiO$_3$ thin film has been epitaxially grown on a (100)$_{\text{pc}}$-oriented LaAlO$_3$ substrate by pulsed laser deposition, see Methods. Subsequently, a PAuM is transferred onto the thin film sample. 
As bulk, LaNiO$_3$ and LaAlO$_3$ crystallize in a rhombohedral perovskite-type structure with the space group characterized by anti-phase rotations of the octahedra, $a^-a^-a^-$ in Glazer’s notation (see Fig.~\ref{fig:LNO1}a). 
The structure gives rise to five Raman-active vibrational modes $\Gamma_{\text{Raman}}$  =  $A_{1\text{g}}$ + 4 $E_{\text{g}}$ \citep{Chaban2010a}. 
Epitaxially strained LaNiO$_3$ on LaAlO$_3$ stabilizes a monoclinic structure with the space group $C2/c$ \cite{Schober2020}. 
However, for simplicity reasons and its close shape of the Raman spectrum, we retain the notations of the rhombohedral bulk symmetry. 
For our Raman spectroscopy experiment, we use a excitation wavelength of \unit[785]{nm}, as it shows the best performance with PAuM, see Table~\ref{TAB:SSR_factorss} and Ref.~\cite{Wyss:2022a} (see Supporting Information S10 for the spectra under \unit[660]{nm} excitation). 
Figure~\ref{fig:LNO1}b shows the Raman spectra of LaNiO$_3$ on LaAlO$_3$ with (top) and without PAuM (bottom). 
The light-blue and gray boxes correspond to regions with vibrational bands of LaAlO$_3$ and LaNiO$_3$, respectively. 
The Raman spectrum measured without PAuM is in perfect agreement with literature data~\cite{Weber2016b, Schober2020}.
Yet, the comparison of the spectra with and without PAuM reveals a number of striking differences:
With the PAuM, the signal of the LaAlO$_3$ substrate is barely visible and can only be approximated as shoulder-like features in the region between \unit[100]{cm$^{-1}$} and \unit[160]{cm$^{-1}$}. 
Furthermore, the LaNiO$_3$-$E_g$ mode around \unit[400]{cm$^{-1}$} exhibits a prominent difference. 
In the spectrum without PAuM, we find a single, albeit asymmetric, peak at \unit[412]{cm$^{-1}$}. 
In contrast, the spectrum with PAuM has two distinct features centered at \unit[389]{cm$^{-1}$} and \unit[413]{cm$^{-1}$}. (For a high resolution Raman spectrum of this region see Supporting Information S10) 
The $A_{1\text{g}}$ mode of LaNiO$_3$, on the other hand, shows only a minor shift of $\sim$ \unit[3]{cm$^{-1}$} from \unit[213]{cm$^{-1}$} without PAuM to \unit[216]{cm$^{-1}$} with PAuM.
Note, at low frequencies, the spectrum with PAuM is characterized by an intensity increase. We assign this increase to the onset of intense low frequency $E_{\text{g}}$-mode of LaNiO$_3$ at \unit[74]{cm$^{-1}$} below the spectral cut-off of the measurement set-up.

Our results allow for a number of interesting conclusions: 
First, in the presence of the PAuM, the thin film signal is strongly enhanced with respect to the LaAlO$_3$ substrate signal.
Second, the $E_{\text{g}}$-mode splitting in the Raman spectrum demonstrates that the PAuM-enhanced Raman signal does not represent an average of the vibrational signal of the entire film. Rather, the PAuM-enhanced Raman signal stems primarily from the surface layer, namely the first few nanometers where the the PAuM enhancement is most effective.

From the material perspective, we know that the crystal structure changes significantly within these first few nanometers \citep{Fowlie2017}. 
A look at the vibrational patterns reveals further details of those surface distortions.
The $A_{1\text{g}}$ and $E_{\text{g}}$ around \unit[213]{cm$^{-1}$} and \unit[403]{cm$^{-1}$} of bulk-like LaNiO$_3$ correspond to an octahedral tilt and bending vibration patterns, respectively \citep{Weber2016b,Chaban2010a}. Therefore, we suggest that deformations of the octahedra, permitted by the $C/2c$ symmetry, dominate the structural changes in the surface region over changes of the octahedron tilt-angles. 

Note that the PAuM plasmonic enhancement of the Raman spectra of oxide materials differs from graphene, our model surface previously, where the primary scattering volume is reduced. Hence, the Raman spectra with a PAuM has a lower total intensity and signal-to-noise ratio than without the PAuM, despite the enhancement of surface Raman signals (Fig.~\ref{fig:LNO1}). Overall, our findings demonstrate that enhancement of the Raman signal by PAuM allows the effective extraction of the Raman response of an oxide surface. This novel spectroscopy-based access to the structure reveals major structural changes at the film surface compared to the bulk of the film, in agreement with the monoclinic symmetry of the film.

\section{Discussion}

The improvement of the surface-to-bulk Raman intensity ratio by up to three orders of magnitude is the primary quality of the PAuM. Moreover, PAuM  
 can sustain excitation power densities up to \unit[$10^6$]{W\,cm$^{-2}$} for \unit[785]{nm} excitation without structural damage. This exceeds the threshold of classical plasmonic structures by two orders of magnitude~\cite{Wyss:2022a} and further increases the detectable Raman signal from a surface. Furthermore, surface-sensitive Raman measurements with PAuM equally function at cryogenic temperatures (see Supporting Information S9). Hence, PAuM enhanced Raman spectroscopy may give access to temperature-driven phase transition of a surface layer\cite{Zhang:2005a}.

The geometrical randomness of the pores in our PAuM always provides pores with suitable enhancement for an arbitrary combination of excitation wavelength, polarization, and refractive index of the material, see Supporting Information S5. We can further tune the ratio between surface enhancement and bulk suppression by altering the PAuM thickness. Thicker PAuM lead to fewer nanopores and reduced transmittance, thinner PAuM to more nanopores and increased transmittance. The alteration of the nanopore geometries as a function of film thickness, furthermore, allows to adjust its resonance. 

\section{Conclusions}
We have demonstrated that transferable nanoporous gold membranes (PAuM) enable surface-sensitive Raman spectroscopy. Nanopores in the membrane act as plasmonic hot-spots for enhanced Raman scattering. Simultaneously, the membrane nature of the PAuM suppresses bulk Raman signals. Using graphene as a model surface, we have shown an increase of the Raman surface-to-bulk ratio of a factor 1100. Simulations combined with Raman measurements on buried graphene samples showed that the enhancement drops exponentially with distance from the PAuM. 90\% of the enhancement occur within the top \unit[2.5]{nm} of the probed material. To demonstrate the ultility of our approach, we applied it to an open scientific question -- the spectroscopic analysis of the surface structure of LaNiO$_3$. By PAuM-enabled surface-sensitive Raman scattering of a LaNiO$_3$ thin film on LaAlO$_3$, we found major structural changes at the surface of LaNiO$_3$, which had not been observed by Raman spectroscopy to date. Surface-sensitive Raman scattering, as introduced in this work, therefore extends the use of Raman spectroscopy as a surface analytical technique. Our approach is not limited to crystalline surfaces, but may also be employed to monitor surface-bound chemical reactions or to characterize biological membranes.

\section{Methods}
\textbf{PAuM Manufacturing and transfer:} A non-continuous gold film of \unit[20]{nm} is evaporated on a silicon/silicon dioxide (Si/SiO2) Wafer with an oxide thickness of 30 nm at a rate of \unit[0.2]{nm/s}. The PAuM is subsequently coated with poly(methyl metacrylate) (PMMA) in Anisol (\unit[2]{w\%}), followed by a floating etch in buffered hydrofluoric acid (BHF), releasing the PAuM/PMMA from the substrate. The PAuM/PMMA is subsequently rinsed by floating on deionised water for a total of \unit[60]{min}, after which it is scooped with the target substrate and let dry. Oxygen plasma (for non-organic samples, \unit[400]{W} for \unit[4]{min}) or acetone/isopropyl alcohol baths (for organic samples) are used to remove the PMMA. (See Supporting Section S1 for details). Freestanding PAuM for pore size characterization and transmission measurement are obtained using a pre-patterned Si/Si$_3$N$_4$ chip with arrays of 64 x \unit[4]{$\mu$m} holes as described by Celebi \textit{et al.} \cite{celebi2014ultimate}.

\textbf{Optical Transmission of PAuM:} Transmission measurements were performed using a self-built setup by focussing white light from broadband supercontinuum laser (NKT) with an objective (NA 0.9) on the freely suspended PAuM from the top. A long-distance objective (NA 0.7) place below the PAuM collected the transmitted light, which was recorded with a Princeton Instruments Acton spectrometer. Reference spectra taken without the PAuM were used to substract the background and eliminate any wavelength-dependence of the spectrometer. 

\textbf{Raman spectroscopy} Raman measurements were performed on a Horiba LabRam Raman spectrometer equipped with a motorized stage. Laser powers were kept below \unit[3]{mW} (100X Objective, NA 0.9) with integration times up to between 1s and 20s.

\textbf{LaNiO$_3$ thin film growth:} The LaNiO$_3$ film was grown on a single crystalline LaAlO$_3$ (001)$_{\text{pc}}$ substrate (CrysTec GmbH) by pulsed laser deposition using a 248\,nm KrF excimer laser (LPXpro, Coherent Ltd.). The film was grown at a substrate temperature of 700$^{\circ}$C under an oxygen partial pressure of 0.1\,mbar. The laser fluence was set to 1\,J/cm$^2$ with a repetition rate of 2\,Hz. X-ray diffraction measurements were performed on a four-circle thin-film diffractometer (PanAlytical X’Pert3 MRD) with Cu K$\alpha1$ radiation ($\lambda$ = 1.54\,\AA). X-ray reflectometry was performed to quantify the LaNiO$_3$ film thicknesses, see Supporting Information S10. Surface topography measurements were conducted using atomic force microscopy in a Bruker Multimode 8 scanning probe microscope with Pt-coated Si tips (MikroMasch, k = 5.4\,N/m).

\section{Author contributions}
M.C.W., M.F., R.M.W, and S.H. conceived the project. R.M.W. and K.P.S. fabricated the nanoporous membranes and graphene samples, and performed the transfers of PAuM onto target substrates, supervised by J.V.. M.P. and S.H measured the optical transmission of the PAuM. S.H. performed and analysed the wavelength-dependent Raman measurements. E.B. and S.H. performed the Raman measurements at cryogenic temperatures. G.K. performed the numerical simulations. M.F.S. and M.T. fabricated the LaNiO$_3$ thin film on LaAlO$_3$. M.C.W. performed and analysed the Raman measurements on the LaNiO$_3$ thin films with support from S.H.. L.H., G.M., M.P., M.F., and L.N. assisted in the Raman measurements and numerical simulations. R.M.W., G.K., M.F., M.C.W. and S.H. interpreted the results and co-wrote the manuscript with input from all authors. S.H. coordinated and supervised the project.

\section{Acknowledgments}
R.M.W. thanks the Binnig and Rohrer Nanotechnology Center in Rueschlikon/Switzerland and ETH Zurich Department of Materials. G.K. was funded by the Deutsche Forschungsgemeinschaft (DFG, German Research Foundation) - Project-ID 182087777 - SFB 951. S.H. acknowledges funding from the Deutsche Forschungsgemeinschaft (DFG) under the Emmy Noether Initiative (Project-ID 433878606) and from financial support by ETH Z\"urich Career Seed Grant SEED-16 17-1. M.T. acknowledges the financial support by the Swiss National Science Foundation under project No. 200021\_188414. M.F. and L.N. acknowledges the financial support by the Swiss National Science Foundation under project No. 200020\_192362. M.C.W. is grateful for financial support by the Région des Pays de la Loire under the Etoile Montante Initiative (2022\_11808) and the PULAR Academy. The authors express their gratitude to Marcela Giraldo for initiating the meeting that started this project. The authors acknowledge the use of the facilities at the Scientific Center for Optical and Electron Microscopy (ScopeM) at ETH Zu\"urich.

\section{Competing financial interest}
RMW, LH, GM and SH are planning to use porous gold membranes commercially. RMW holds a patent on manufacturing porous gold membranes (Patent US17/294748) Otherwise, there are not competing financial interests.

\section{References}
\bibliographystyle{MSP}
\bibliography{Bib_SSR_Paper_v1}

\end{document}